\documentclass[journal,onecolumn]{IEEEtran}

\makeatletter
\def\ps@headings{%
\def\@oddhead{\mbox{}\scriptsize\rightmark \hfil \thepage}%
\def\@evenhead{\scriptsize\thepage \hfil \leftmark\mbox{}}%
\def\@oddfoot{}%
\def\@evenfoot{}}

\usepackage{amsmath}
\usepackage{colordvi}
\usepackage{amsfonts}
\usepackage{graphicx}
\usepackage{epsfig}
\usepackage{color}
\usepackage{url}
\usepackage{multirow}

\def\conv{*}

\def\eps{\varepsilon}
\def\P{\mathbb{P}}

\def\R{\mathcal{R}}

\def\et{\textit{et al.}}
\newcommand{\argmax}{\operatornamewithlimits{argmax}}

\newtheorem{theorem}{Theorem}

\newtheorem{lemma}{Lemma}

\begin{document}
%
\title{On the Catalyzing Effect of Randomness on the Per-Flow Throughput in Wireless Networks}

\author{Florin Ciucu, Jens Schmitt

\thanks{Florin Ciucu is with Telekom Innovation Laboratories / TU Berlin, Germany, e-mail: florin@net.t-labs.tu-berlin.de.}
\thanks{Jens Schmitt is with the Department of Computer Science, University of Kaiserslautern, Germany, e-mail: jschmitt@cs.uni-kl.de}
}

\maketitle

\begin{abstract}
This paper investigates the throughput capacity of a flow crossing
a multi-hop wireless network, whose geometry is characterized by
general randomness laws including Uniform, Poisson, Heavy-Tailed
distributions for both the nodes' densities and the number of
hops. The key contribution is to demonstrate \textit{how} the
\textit{per-flow throughput} depends on the distribution of 1) the
number of nodes $N_j$ inside hops' interference sets, 2) the
number of hops $K$, and 3) the degree of spatial correlations. The
randomness in both $N_j$'s and $K$ is advantageous, i.e., it can
yield larger scalings (as large as $\Theta(n)$) than in non-random
settings. An interesting consequence is that the per-flow capacity
can exhibit the opposite behavior to the network capacity, which
was shown to suffer from a logarithmic decrease in the presence of
randomness. In turn, spatial correlations along the end-to-end
path are detrimental by a logarithmic term.
\end{abstract}


%

\section{Introduction}
A groundbreaking work at the intersection between communication
networks and information theory is a set of network capacity
results obtained by Gupta and Kumar~\cite{Gupta00}. Under some
simplifications at the network layers (e.g., no multi-user coding
schemes, or ideal assumptions on power control, routing, and
scheduling), those results establish asymptotic scaling laws on
the maximal data rates which can be reliably sustained in
multi-hop wireless networks. They have further inspired a
tremendous research interest and provided fundamental insight into
the development of the long desirable functional network
information theory (see Andrews~\et~\cite{Andrews08commag}).

A major assumption in the network model from~\cite{Gupta00}, which
was widely adopted thereafter, is that the geometry follows a
\textit{uniform} distribution. This assumption implies that the
nodes' interference sets follow the binomial (or its Poisson
approximation) distribution. A small set of works considered
geometries with specific heterogeneous geometries, which were
shown to play a fundamental role in the capacity scaling laws~(see
the Related Work Section). The goal of this paper is to go beyond
specific random geometries, by analyzing (throughput) capacity
results in networks with \textit{general}
distributions\footnote{In this paper, the notions of throughput
and (throughput) capacity, used interchangeably, explicitly refer
to the maximal flows' throughput achieved in some network, with no
coding schemes being considered. The meaning of \textit{capacity}
employed herein differs thus from the one in information theory.}.
To make the analysis manageable, the paper assumes that network
nodes implement the Aloha MAC protocol; the simplicity of the
protocol permits the derivation of per-flow capacity results in
\textit{closed-form} and \textit{explicit} up to the optimization
of a single parameter.

We point out that the paper particularly focuses on
\textit{per-flow capacity} results, rather than the network
capacity results which are mostly sought in the literature. The
advantage of per-flow capacity results is that they determine the
network capacity by a summation argument; in contrast, a division
argument to compute the per-flow capacity from the network
capacity only holds for uniform geometries. Moreover, unlike the
majority of existing \textit{asymptotic} capacity results---whose
practicality is often questioned for small to medium sized
networks (Akyildiz and Wang~\cite{Akyildiz09}, p. 180)---this
paper provides \textit{non-asymptotic} results, e.g., the per-flow
capacity in (finite) $T$ time units in a network with (finite) $N$
nodes.

Intuitively, different randomness laws in the geometry yield
different (per-flow) capacity results. This paper goes beyond this
simple intuition and makes three key contributions:
\begin{enumerate} \item Analyzing the beneficial role of
randomness in the network's geometry on the per-flow
capacity.\item Quantifying the magnitude of the benefit, referred
to as \textit{randomness gain}, in terms of scaling laws. \item
Providing the `best' distributions maximizing the randomness
gain.\end{enumerate}

To properly summarize our observations, let us briefly describe
the network model; see Section~\ref{sec:model} for the complete
description. A flow crosses some end-to-end (e2e) path with $K$
hops. The interference set of a hop $j$ consists of $N_j$
(interfering) nodes, which is referred to as the hop density; all
$N_j$'s and $K$ can have general distributions. The degree of
spatial correlations, denoted by $\gamma$, defines the number of
consecutive hops with correlated densities; for instance, in the
practical scenario when all $N_1,~N_2,\dots,N_K$ are statistically
correlated, then $\gamma:=K$.

The closed-form expressions of the derived capacity results allow
a sensitivity study of the various parameters in the network
model, e.g., $N_j$'s, $K$, and $\gamma$. From this study we
collected the following observations:
\begin{enumerate}
\item[O1.] By scaling $N$ (here a shorthand for $N_j$'s), the
per-flow capacity depends on $N$'s distribution under a
sample-path neighbor-aware probabilities assumption. Concretely,
nodes should set transmission probabilities, explicitly or
implicitly, according to the nodes' densities (transmission
probabilities should be roughly proportional to the number of
neighbors). With this assumption, different distributions of $N$
with identical averages can yield different scaling laws. In turn,
when nodes use a fixed optimized transmission probability on all
sample paths, the per-flow capacity is sensitive to the
distribution of $N$, but only through its mean $E[N]$.

\item[O2.] By scaling time, the per-flow capacity is invariant to
$K$. In other words, if the system runs over a sufficiently long
time scale, then the per-flow capacity, with the interpretation of
a \textit{rate}, stabilizes and does not depend on the actual
number of hops. More interestingly, by simultaneously scaling both
time and $K$, we find that randomness in $K$ has also a
fundamental impact on the per-flow capacity (i.e., it changes the
order of growth).

\item[O3.] The size of the randomness gain can be as large as
${\mathcal{O}}(n)$ in $N$ and much smaller in $K$, which indicates
that temporal correlations due to $N$ are much more sensitive to
randomness than spatial correlations due to $K$\footnote{By
\textit{temporal correlations} we mainly refer to dependent events
which can occur simultaneously (i.e., transmissions within the
same interference set). By \textit{spatial correlations} we mainly
refer to dependent events which can occur at different times
(i.e., transmissions at two relays whose interference sets share
some nodes).}, as an effect of spatial reuse. Based on specific
distributions of $N$ and $K$, we find the surprising fact that the
randomness gain in $K$ is the logarithm of the randomness gain in
$N$. To clarify its precise meaning, the randomness gain is
defined as the relative difference of the per-flow capacities in
two scenarios: one in which a network parameter (say $N$) is
random, and another in which the same parameter $N$ is set to its
(non-random) expectation $E[N]$.

\item[O4.] By simultaneously scaling time, $K$, and $\gamma$, the
per-flow capacity decays logarithmically in $\gamma$. The
simultaneous scaling is needed as in 2); otherwise, the impact of
spatial correlation vanishes. The observed logarithmic detrimental
factor of spatial correlations, is analogous to an existing result
characteristic to \textit{wired} networks: in a tandem of nodes in
which exponentially sized packets arrive as a Poisson process, the
e2e delay scales as $\Theta(n\log{n})$ and not as
$\Theta(n)$~\cite{BuLiCi11}. The $\Theta(n)$ holds under the
so-called Kleinrock's Independence Assumption that packets
independently regenerate their sizes at each hop. Without this
assumption, the extra logarithmic factor stems from the spatial
correlations due to the `very large' packets inducing long delays
to packets behind them, and at \textit{each} node. Similarly, in
our settings, the logarithmic term arises when spatial
correlations span across the entire e2e path.
\end{enumerate}

These insights complement some fundamental existing ones. One is
that randomness can have a detrimental role in the network
capacity (see Gupta and Kumar~\cite{Gupta00} or
Franceschetti~\et~\cite{Franceschetti07}). In contrast, our
results show that, in networks with non-uniform geometries,
randomness has a beneficial role in the per-flow capacity; for a
discussion clarifying the apparently contradicting detrimental and
beneficial roles of randomness see the end of
Section~\ref{sec:impactneighbors}. Another important known fact is
that TDMA and CSMA with properly tuned parameters achieve the same
capacity (Chau~\et~\cite{Chau09}). Such a MAC
\textit{insensitivity} result, however, depends on the assumed
uniform geometry. In turn, our results indicate that in a
single-hop scenario with non-uniform random geometry, the per-flow
capacity is in fact \textit{sensitive} to the MAC since CSMA
implicitly satisfies the sample-path assumption, whereas explicit
overhead would be needed for Aloha or TDMA.

The rest of the paper is organized as follows. First we discuss
related work. In Section~\ref{sec:model} we introduce the network
model, and in Section~\ref{sec:tools} we introduce the main
analytical tools enabling the capacity analysis. In
Section~\ref{sec:capacity} we first present the main result of the
paper, i.e., non-asymptotic bounds on the capacity of a fixed
source-destination pair, and then we investigate the capacity's
sensitivity to the randomness factors in geometry. Brief
conclusions are presented in Section~\ref{sec:conclusions}.

\section{Related Work} Gupta and Kumar~\cite{Gupta00} analyzed
the asymptotic capacity of homogeneous random networks with
uniformly distributed nodes, and showed the notorious
$\Theta\left(1/\sqrt{n\log{n}}\right)$ scaling law on the per-flow
capacity under a specific communication channel model. This law
was improved to $\Theta\left(1/\sqrt{n}\right)$ for another
channel model by Franceschetti~\et~\cite{Franceschetti07}. Under a
mobility model and a two-hop relay model, the per-flow scaling
laws were further improved to $\Theta(1)$, i.e., the best
achievable one, but at the expense of conceivably long delays
(Grossglauser and Tse~\cite{Grossglauser02}). For a more
comprehensive review of related scaling laws see Xue and
Kumar~\cite{Xue06}.

Asymptotic capacities for heterogeneous networks, e.g., not
necessarily with uniformly distributed nodes, were derived in
special cases. Toumpis~\cite{Toumpis04} considered a logically
clustered network in which $n$ sources communicate with $n^d$
cluster heads (yet all are uniformly placed), and showed that
network capacity degrades in the presence of bottlenecks when
$0<d<0.5$. Perevalov~\et~\cite{Perevalov06} considered physically
clustered networks, with uniformly placed nodes and clusters of
nodes, and showed that network capacity fundamentally depends on
the size of the clusters. For some clustered networks, Kulkarni
and Viswanath~\cite{Kulkarni04} showed that network capacity
preserves the scaling law from~\cite{Gupta00}. For some other
specific clustered networks, however, Alfano~\et~\cite{Alfano10}
and Martina~\et~\cite{Martina10} recently showed that the per-flow
capacity is fundamentally influenced by the geometry. Similar
results have also been reported from simulations by
Hoydis~\et~\cite{HoydisPM09}. Our paper differs from these works
in that it provides per-flow and (non)-asymptotic capacity results
for a broad range of random geometries.

As far as non-asymptotic capacity results are concerned, many
exist in the single-hop case (e.g., Kleinrock and
Tobagi~\cite{KleinrockTob75} or Bianchi~\cite{Bianchi00}); the
latter is derived for 802.11 DCF networks, by assuming that all
nodes independently see the system in steady state. Much fewer
results exist in multi-hop networks, mostly under simplifying
technical assumptions and approximations to deal with the
intrinsically hard problem of spatio-temporal correlations.
Capacity results were computed in 802.11 DCF networks, modelled as
contention graphs, under the assumption that collision
probabilities are mainly due to hidden node interference
(Gao~\et~\cite{GaoCL06}). E2e delays in both TDMA and Aloha
line-networks were investigated using a decomposition approach,
which relies on the approximation that the departure processes at
the relay nodes have independent inter-departure times (Xie and
Haenggi~\cite{XieH09}). The delay analysis of wireless channels
under Markovian assumptions was studied by
Zheng~\et~\cite{Zheng13}, and the delay analysis of multi-hop
fading channels by Al-Zubaidy~\et~\cite{Zubaidy13}.

Closer to our work, non-asymptotic per-flow capacity bounds were
derived in networks with \textit{non-random} $N_j$'s and $K$, and
\textit{no} spatial correlations (i.e., $\gamma=1$) (see
Ciucu~\et~\cite{CiHoHu10,CiucuISIT11,CiKhJiYaCu13}). This paper
extends these results by accounting for general randomness in
$N_j$'s, $K$, and spatial correlations (i.e., $\gamma>1$).

Our results on the advantageous effect of randomness relate to a
``folk theorem" from queueing theory which states that, when the
mean inter-arrival (service) time is fixed, the constant
inter-arrival (service) time distribution \textit{minimizes}
queueing metrics such as average waiting time. Such results were
proven for renewal processes (Rogozin~\cite{Rogozin66}) and also
for more general arrival processes with exponential service times
(Hajek~\cite{Hajek83} and Humblet~\cite{Humblet82}). Moreover, our
results on the bimodal nature of distributions maximizing the
randomness gain agree with parallel results from queueing theory.
For instance, bimodal distributions maximize queue lengths in
GI/M/1 queues (Whitt~\cite{Whitt84a}), in G/M/1 queues with bulk
arrivals (Lee and Tsitsiklis~\cite{Lee92}), or in queues with bulk
arrivals and finite buffers (Bu\v{s}i\'{c}~\et~\cite{Busic06}).

\section{Network Model}\label{sec:model} In this section we
describe the network model and the type of capacity results
investigated in this paper.

\begin{figure}[t]
  \centering
   \includegraphics[scale=0.57]{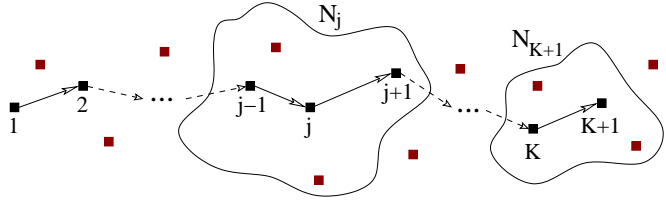}
   \caption{A multi-hop wireless network with a random number of $K$ hops. The interference set (IS) of node $j$ contains a random number $N_j$ of nodes. The performance metric of interest is the (non-)asymptotic e2e capacity of node $1$ transmitting to
node $K+1$ using the nodes $2,3,\dots,K$ as
relays.}\label{fig:multinode}
  \end{figure}

We consider a general network model accounting for three
randomness sources, thus significantly generalizing related
models. Concretely, we consider the multi-hop random network
geometry from Fig.~\ref{fig:multinode}. Node $1$ (the source)
transmits to node $K+1$ (the destination) using nodes
$2,3,\dots,K$ as relays, where $K$ is a random variable denoting
the number of hops. The number of nodes inside the interference
set (IS) of node $j$, and \textit{excluding} node $j$, is denoted
by the random variable $N_j$, for $j=2,3,\dots,K+1$; $N_j$'s are
also referred to as nodes' (hops') densities. The ISes allow to
model arbitrary interference models and do not rely on geometrical
assumptions like disc-based transmission or interference ranges.

One requirement is the existence of an e2e path between the source
and the destination. This assumption is motivated by the very goal
of the paper, i.e., the derivation of per-flow capacities which
requires the flow to be well-defined in terms of an e2e path. If
such e2e paths were subject to discontinuities, the derived
capacity results would still hold for the transient regimes during
which an e2e path exists.

The model further needs knowledge of the distributions of
$N_{j}$'s and $K$, which characterize the first two randomness
sources. These distributions can be quite general; in fact, the
capacity formulae from the main result (see
Theorem~\ref{th:e2ecap}) allow plugging-in any specific
distribution law in order to quantify the underlying impact.

Concretely, all $N_j$'s are finite and identically distributed
with density
\begin{equation}
\pi_n=\P\left(N=n\right),~n=2,\dots,n_{\textrm{max}}~.\label{eq:densityN}
\end{equation}
$N$ generically stands for $N_j$'s. Note that $\pi_1=0$, i.e., the
nodes on the e2e path are not isolated, and $n_{\textrm{max}}$
denotes the maximum number of nodes inside an IS. The assumption
of identically distributed $N_j$'s is a mild one and mitigates the
notational complexity.

For the second randomness source, we assume that the number of
hops $K$ is finite, statistically independent of all the $N_j$'s,
and has the density
\begin{equation}
\tilde{\pi}_{k}=\P\left(K=k\right)~,k=1,2,\dots,k_{\textrm{max}}~.\label{eq:densityK}
\end{equation}
The maximal number of hops is $k_{\textrm{max}}$. The independence
assumption simplifies the proof of the main result, i.e.,
Theorem~\ref{th:e2ecap}; the proof could also account for the
conditional distributions $\P\left(N_j=n\mid K=k\right)$ at the
expense however of increasing the notational complexity.
Nevertheless, the independence assumption between $K$ and $N_j$'s
is conceivably strong in small networks. Indeed, if nodes set
large interfering ranges then the $N_j$'s are also large and $K$
is small; in turn, if the interfering ranges are small then the
$N_j$'s are also small and $K$ is large. Such correlation effects
lessen in large network regimes, whereby our later asymptotic
analysis applies.

The third source of randomness in the network model concerns the
degree of spatial correlations, i.e., the degree of statistical
dependence amongst $N_j$'s. Dealing with all possible combinations
and types of such dependencies is clearly an overwhelming task. In
our analysis we assume that for each $j=1,2,\dots,k_{max}$, the
r.v. $N_j$ is statistically independent of all $N_i$'s with
$i\in\left\{j+\gamma,j+\gamma+1,\dots,k_{max}\right\}$. By the
commutativity of the independence relationship between two r.v.'s,
$N_j$ is also statistically independent of $N_i$'s with
$i\in\left\{1,2,\dots,j-\gamma\right\}$.

This particular \textit{dependency parameter} $\gamma$
characterizes the maximal number of consecutive ISes for which
dependencies (correlations) may exist between the first and the
rest. For instance, if $\gamma=1$ then all $N_j$'s are
statistically independent; at the other extreme, in a static
scenario with $k$ hops, $\gamma=k$ corresponds to dependencies
between any pair of $N_1,~N_2,\dots,~N_k$ (this latter scenario is
conceivably the \textit{most practical} one).

Besides the description of the three randomness sources, the
network model requires the specification of the MAC protocol.
Concretely, given a slotted time model, the network model requires
that all nodes transmit with some probability $p$, independently
of each other, and from slot to slot. This requirement is
immediately satisfied by the slotted-Aloha protocol
(Abramson~\cite{Abramson70}), and implicitly satisfied by 802.11
DCF under an independence assumption
from~Bianchi~\cite{Bianchi00}, in the single-hop case. The
assumption from~\cite{Bianchi00}, which is commonly adopted in the
literature, states that all nodes see the channel in steady state,
while the transmission probability $p$ can be computed from the
parameters of the protocol (see Eqs.~(7) and (9)
in~\cite{Bianchi00}). An additional approximation for 802.11 DCF
is the consideration of an average slot size (given also
in~\cite{Bianchi00}, in Eq.~(13)). In a multi-hop case the
steady-state assumption becomes however less justifiable, due to
the hidden node problem, and further assumptions are needed
(Gao~\et~\cite{GaoCL06}). Since we are seeking rigorous capacity
results, in order to perform a rigorous analysis of the underlying
roles of the three randomness sources, we mainly adopt
slotted-Aloha and make side remarks on 802.11 DCF and also on
TDMA.

Denoting the transmission from node $i$ to $j$ by $[i\rightarrow
j]$, a transmission $[j\rightarrow j+1]$ is successful if node $j$
is the only transmitting node inside the IS of node $j+1$, in a
time slot. As far as data sources are concerned, we assume that
the nodes $2,3,\dots,K$ have infinite buffers and only relay the
data from node $1$ to node $K+1$. Also, all the other nodes are
saturated, i.e., they always attempt to transmit according to the
MAC protocol. This saturation assumption implies that the computed
capacity of the path $[1\rightarrow K+1]$ is conservative from the
data-link layer perspective.

The concrete type of capacity investigated in this paper is the
per-flow (non-)asymptotic throughput capacity of the transmission
$[1\rightarrow K+1]$. Denote by $A(t)$ the arrival process at node
$1$ (i.e., counting the data units to be transmitted to node
$K+1$). Also, denote by $D(t)$ the corresponding departure process
at node $K+1$ (i.e., counting the data units arrived from node $1$
up to time $t$) . For some fixed violation probability $\eps$, a
probabilistic upper bound on the (e2e throughput) capacity rate is
a value $\lambda^U_t$ such that
\begin{equation}
\P\Big(D(t)\geq \lambda_t^U t\Big)\leq\eps~.\label{eq:upperbound}
\end{equation}
In turn, a probabilistic lower bound on the capacity rate is a
value $\lambda^L_t$ such that
\begin{equation}
\P\Big(D(t)\leq\lambda_t^L t\Big)\leq\eps~.\label{eq:lowerbound}
\end{equation}

We point out that the derived capacity rates $\lambda_t^U$ and
$\lambda_t^L$ are obtained in transient (non-asymptotic in the
time scale) regimes, whereas the asymptotic results are
immediately obtained by letting $t\rightarrow\infty$. The other
non-asymptotic regime is in the nodes' densities $N$ and number of
hops $K$; again, related asymptotic results follow directly by
taking limits.

\section{Analytical Tools}\label{sec:tools}
In this section we introduce the modelling tools for the per-flow
capacity problem. The main engine behind the derivations in the
paper is the framework of the stochastic network calculus, in
particular following Fidler~\cite{Fidler06}. The advantage of the
calculus approach is that it considerably simplifies the
complexity of modelling a whole network by (logically) reducing it
to a single-hop only. It is thus sufficient to model any
single-hop scenario from Fig.~\ref{fig:multinode} with source $j$
and destination $j+1$; the multi-hop model will directly follow
from the convolution theorem in the stochastic network calculus.

Let us model an arbitrary single-hop where the source is node $j$.
For each node $l$ inside the IS of node $j+1$ we associate a
random process $X_l(t)$, where $t$ represents time. For every $l$
and $t$, $X_l(t)$ is a Bernoulli random variable (r.v.) taking
value $1$ with probability $p$; with abuse of notation, $l=1$
refers to the source $j$. These r.v.'s are mutually
\textit{i.i.d.} in both time and space, and conditioned on the
realizations of the number of nodes/hops (for a parallel
analytical framework, dealing with non-necessarily independent
increments of $X_l(t)$, e.g., when modelling CSMA/CA besides
Aloha, we refer to Ciucu~\et~\cite{CiKhJiYaCu13}).

Next we introduce the key process for computing the per-flow
capacity. This is referred to as the (virtual) interfering process
of the transmission $[j\rightarrow j+1]$, and is defined through
its increments $V(t-1,t):=V(t)-V(t-1)$ for $t\geq1$ as
\begin{equation}
V(t-1,t)=1-X_1(t)\prod_{l=2}^{N_{j+1}}\left(1-X_l(t)\right)~.\label{eq:vip}
\end{equation}
The initial value is $V(0)=0$. We make the important remark that
$V(t)$ does not depend on whether the source $j$ is saturated or
bursty, since it is defined independently of the arrival process
$A(t)$ at the source. Moreover, as we mentioned earlier, our
analysis handles the situation of idle periods characteristic to
relay nodes $j\geq 2$, and which is due to internal burstiness:
there is nothing to transmit at some slot, and yet the MAC
protocol may successfully select the relay node to access the
channel. Due to such situations we emphasize the attribute
\textit{virtual} for the process $V(t)$.

Next we obtain the moment generating function (MGF) and the
Laplace transform of $V(t)$, needed to derive the upper and lower
bounds, respectively, from Eqs.~(\ref{eq:upperbound}) and
(\ref{eq:lowerbound}). For some parameter $\theta>0$, these
transforms are defined as
\begin{equation*}
M_t(\theta)=E\left[e^{\theta
V(t)}\right]~\textrm{and}~L_{t}(\theta)=E\left[e^{-\theta
V(t)}\right]~.
\end{equation*}
The MGF follows from the backwards equations using conditioning,
and also using the independent increments property of $V(t)$,
i.e.,
\begin{eqnarray*} M_{t+1}(\theta)&=&M_{t}(\theta)E\left[e^{\theta V(1)}\right]\notag\\
&=&M_t(\theta)\sum_iE\left[e^{\theta V(1)}\mid
N(0)=i\right]\P\left(N(0)=i\right)\notag\\
&=&M_t(\theta)\sum_i\left(e^{\theta}q_i+1-q_i\right)\pi_i\notag~.
\end{eqnarray*}
Therefore,
\begin{eqnarray}
M_{t}(\theta)= b_{\theta}^t~,\label{eq:mgfresult}\end{eqnarray}
where $$b_{\theta}=1+q\left(e^{\theta}-1\right),~q=\sum_{l\geq
2}q_l\pi_l,~q_l=1-p(1-p)^{l-1}~,$$ and where the $\pi_l$'s are
from Eq.~(\ref{eq:densityN}). The Laplace transform follows by a
sign change, i.e., $$L_{t}(\theta)=b_{-\theta}^t~.$$

The critical role of the virtual process is to link the arrival
process $A(t)$ at the source $j$ with the corresponding departure
process $D(t)$ at the destination $j+1$. This relationship is
expressed in terms of a stochastic service process, which is a key
tool in the stochastic network calculus (Chang~\cite{Book-Chang},
Jiang and Liu~\cite{Jiang08}), and which is instrumental herein
for the derivation of e2e capacity results. The next Lemma from
Ciucu~\cite{CiucuISIT11} formally establishes this relationship.

\begin{lemma}{({\sc{Single-Hop (Exact) Service Representation}})}\label{th:shsc}
Consider the interfering process $V(t)$ from Eq.~(\ref{eq:vip}).
Then the bivariate random process
\begin{equation}
S(s,t)=t-s-V(s,t)\label{eq:scexpr}
\end{equation}
is an \textit{exact} stochastic service process for node $A$,
i.e.,
\begin{equation}
D(t)=A\conv S(t)~\textrm{a.s.}~,\label{eq:esc}
\end{equation}
for all arrival processes $A(t)$. Here, the symbol `$*$' denotes
the $(\textrm{min},+)$ convolution operator defined for all
$t\geq0$ as $A\conv S(t):=\inf_{0\leq s\leq
t}\left\{A(s)+S(s,t)\right\}~.$
\end{lemma}

The process $S(s,t)$ quantifies the service received over the link
$[j\rightarrow j+1]$. A key observation is that Eq.~(\ref{eq:esc})
holds for \textit{all} arrival processes $A(t)$. This invariance
is instrumental for carrying out the incoming multi-hop analysis,
by circumventing the intrinsically difficult (queueing) problem
that the arrival processes at the relay nodes $j\geq2$ are hard to
characterize. The process $D(t)$ from Eq.~(\ref{eq:esc}) can be
also viewed as the output from a variable capacity node~(Boudec
and Thiran~\cite{Book-LeBoudec}), given the definition of the
interfering process $V(t)$ from Eq.~(\ref{eq:vip}). Moreover,
$S(s,s+1)$ can be viewed as the instantaneous per-flow
\textit{effective capacity}, as proposed to model the
instantaneous channel capacity by Wu and Negi~\cite{WuNegi03}. In
turn, the process $V(s,t)$ can be viewed as an \textit{impairment
process} as defined by Jiang and Liu~\cite{Jiang08}, p. 72. Having
available an exact service process for single-hop transmissions,
we can next derive both upper and lower bounds on e2e capacity.

\section{End-to-End Per-Flow Capacity}\label{sec:capacity} In this section we first derive the main result of this paper,
i.e., a closed-form expression for the per-flow capacity along the
e2e path from Fig.~\ref{fig:multinode}. Then we investigate its
sensitivity to the three randomness sources in the network model:
the distribution of the number of neighbors $N_j$'s and the number
of hops $K$, and the dependency parameter $\gamma$.

The procedure for getting the upper and lower capacity bounds
follows the methodology of the stochastic network calculus (see
Boudec and Thiran~\cite{Book-LeBoudec}, Chang~\cite{Book-Chang},
and Jiang and Liu~\cite{Jiang08}). First, service processes
$S_j(s,t)$ for each single-hop transmission $[j\rightarrow j+1]$,
$j=1,2,\dots,K$, are constructed as in Eq.~(\ref{eq:scexpr}).
These processes are then convolved in the underlying
$(\textrm{min},+)$-algebra yielding the (network) service process
\begin{equation}
S(s,t):=S_1\conv S_2\conv\dots\conv S_K(s,t)~,\label{eq:e2esc}
\end{equation}
which characterizes the available service (or the per-flow
effective capacity in the terminology from Wu and
Negi~\cite{WuNegi03}) along the e2e path $[1\rightarrow K+1]$.
Eq.~(\ref{eq:e2esc}) is the convolution theorem from network
calculus which reduces the multi-hop analysis to a single-hop
analysis. The advantage of the theorem is that it circumvents the
difficult problem of modelling input/output processes at
intermediate relay nodes, as mentioned earlier.

\begin{theorem}{({\sc{Non-Asymptotic Capacity Bounds}})}\label{th:e2ecap}
Consider the multi-hop network model from Section~\ref{sec:model}
with dependency parameter $\gamma$. Let $q_l=1-p(1-p)^{l-1}$,
$q=\sum_{l\geq 2}q_l\pi_l$, and
$b_{\theta}=1+q\left(e^{\theta}-1\right)$ for any $\theta>0$ (see
Eq.~(\ref{eq:mgfresult})). Then, for some violation probability
$\eps$, a probabilistic lower bound on the e2e capacity is for all
$t\geq k_{max}$
\begin{equation}
\lambda_t^L=\sup_{\theta>0}\left\{1-\frac{1}{\gamma}\frac{\log{b_{\gamma\theta}}}{\theta}+\frac{\log{\eps}}{t\theta}-\frac{c_K}{t\theta}\right\}~,\label{eq:lb}
\end{equation}
where
$c_K=\sum_{k=1}^{k_{\textrm{max}}}\tilde{\pi}_k\log{\binom{t+k-1}{k-1}}$.
The upper bound is
\begin{equation}
\lambda^U_t=\inf_{\theta>0}\left\{1+\frac{1}{\gamma}\frac{\log{b_{-\gamma\theta}}}{\theta}-\frac{\log{\eps}}{t\theta}\right\}~.\label{eq:ub}
\end{equation}
\end{theorem}

We remark that the asymptotic lower and upper bounds coincide
(after using Stirling's approximation for the factorial in the
binomial term, with $\theta=\Theta\left(t^{-\zeta}\right)$,
$0<\zeta<1$), i.e.,
\begin{equation}
\lambda:=\lim_{t\rightarrow\infty}\lambda_t^L=\lim_{t\rightarrow\infty}\lambda_t^U=1-q~.\label{eq:ascap}
\end{equation}
We denote the asymptotic capacity by $\lambda$.
Theorem~\ref{th:e2ecap} generalizes existing non-asymptotic
lower-bound results
from~Ciucu~\et~\cite{CiHoHu10,CiucuISIT11,CiKhJiYaCu13} which hold
for \textit{non-random} $N_j$'s and $K$, and $\gamma=1$. Moreover,
Theorem~\ref{th:e2ecap} also provides the corresponding upper
bounds. We also point out that the results are explicit up to
optimizing after $\theta>0$.


{\sc Proof.}~Let $\P_k$ denote the underlying probability measure
conditioned on $K=k$. Let $t\geq0$ and the service processes
$S_j(s,t)$ for each transmission $[j\rightarrow j+1]$ for
$j=1,2,\dots,k$, as in Eq.~(\ref{eq:scexpr}). Applying the e2e
service process from Eq.~(\ref{eq:e2esc}), we can write
\begin{eqnarray}
\P_k\left(D(t)\leq \lambda_t^L t\right)&\leq&\P_k\left(A\conv
S(t)\leq
\lambda_t^L t\right)\notag\\
&=&\P_k\left(S_1\conv S_2\conv\dots\conv S_k(t)\leq\lambda_t^L
t\right)~,\label{eq:deriv1ststep}
\end{eqnarray}
because of the saturation condition $A(1)=\infty$. Letting $u_0=0$
and $u_k=t$ we can continue above as follows
\begin{eqnarray}
&&\hspace{-1.1cm}\P_k\left(\inf_{0\leq u_1\leq\dots\leq
u_{k-1}\leq t}\sum_{j=1}^k
S_j\left(u_{j-1},u_j\right)\leq \lambda_t^L t\right)\notag\\
&&\hspace{-0.6cm}=\P_k\Bigg(\inf_{0\leq u_1\leq\dots\leq
u_{k-1}\leq
t}\sum_{j=1}^k\Big(u_j-u_{j-1}-V_j\left(u_{j-1},u_j\right)\Big)\leq
\lambda_t^L t\Bigg)\notag\\&&\hspace{-0.6cm}=\P_k\Big(\sup_{0\leq
u_1\leq\dots\leq u_{k-1}\leq t}\sum_{j=1}^k\Big(
V_j\left(u_{j-1},u_j\right)-\left(u_j-u_{j-1}\right)\Big)\geq
-\lambda_t^L t\Bigg)\label{eq:derive2e3}~,
\end{eqnarray}
where $V_j(u_{j-1},u_j)=u_j-u_{j-1}-S_j(u_{j-1},u_j)$. Next, by
applying the Union bound\footnote{For some probability events $E$
and $F$, the union bound states that $\P\left(E\cup
F\right)\leq\P(E)+\P(F)$.} and the Chernoff bound\footnote{For
some r.v. $X$ and $x,\theta\in\R$, the Chernoff bound states that
$\P(X>x)\leq M_{X}(\theta)e^{-\theta x}$.}, we can bound the last
term in Eq.~(\ref{eq:derive2e3}) by
\begin{eqnarray*}
&&\hspace{-0.8cm}\sum_{0\leq u_1\leq\dots\leq u_{k-1}\leq t}
E\left[\prod_{j=1}^ke^{\theta
V_{j}(u_{j-1},u_j)}\right]e^{-\theta\left(1- \lambda_t^L\right)
t}~.\end{eqnarray*} At this point we rearrange the terms in the
product in the expectation as
\begin{eqnarray}
\prod_{j=1}^ke^{\theta
V_{j}(u_{j-1},u_j)}=\prod_{l=1}^{\gamma}\prod_{i\geq0}e^{\theta
V_{l+i\gamma}\left(u_{l+i\gamma-1},u_{l+i\gamma}\right)}~.\label{eq:arrange}
\end{eqnarray}
The terms in the second product in the right-hand side term are
statistically independent. This is true according to the
definition of the dependency parameter $\gamma$, and also because
the underlying Bernoulli r.v.'s in $V_{l+i\gamma}$'s have
independent increments; note the non-overlapping intervals of
$V_{l+i\gamma}$'s. With this observation we can bound the last
expectation by
\begin{eqnarray*}
\left(\prod_{l=1}^\gamma E\left[\prod_{i\geq0}e^{\gamma\theta
V_{l+i\gamma}\left(u_{l+i\gamma-1},u_{l+i\gamma}\right)}\right]\right)^{\frac{1}{\gamma}}=\left(b_{\gamma\theta}^t\right)^{\frac{1}{\gamma}}~,
\end{eqnarray*}
by using H\"{o}lder's inequality, 
where $b_{\gamma\theta}$ has the expression from the theorem with
$\theta$ replaced by $\gamma\theta$. Recall also the MGF of $V(t)$
from Eq.~(\ref{eq:mgfresult}).

Collecting terms we obtain
\begin{eqnarray*}
\P_k\left(D(t)\leq \lambda_t^L t\right)&\leq&\sum_{0\leq
u_1\leq\dots\leq u_{k-1}\leq
t}\left(b_{\gamma\theta}^t\right)^{\frac{1}{\gamma}}e^{-\theta\left(1-\lambda_t^L\right) t}\\
&=&\binom{t+k-1}{k-1}\left(b_{\gamma\theta}^t\right)^{\frac{1}{\gamma}}e^{-\theta\left(1-\lambda_t^L\right)t}~,
\end{eqnarray*}
where the binomial term is the number of combinations with
repetition.

For the upper bound, we recall from Theorem~\ref{th:shsc} that the
service processes $S_j(s,t)$ are exact, and therefore the e2e
service process from Eq.~(\ref{eq:e2esc}) is exact as well; this
property is critical for proving the upper bound. Using again that
$A(1)=\infty$ we can write
\begin{eqnarray}
&&\hspace{-1cm}\P_k\left(D(t)\geq \lambda^U_t
t\right)=\P_k\left(A\conv S(t)\geq
\lambda^U_t t\right)\notag\\
&&=\P_k\left(S_1\conv S_2\conv\dots\conv S_k(t)\geq\lambda^U_t
t\right)\notag\\ &&\leq\inf_{u_1\leq\dots\leq
u_k}\P_k\left(\sum_{j=1}^k S_j(u_{j-1},u_j)\geq\lambda^U_t
t\right)~.\label{eq:upbound2}
\end{eqnarray}
For $u_1\leq\dots\leq u_k$ we can expand the probabilities as
\begin{eqnarray*}
&&\hspace{-1cm}\P_k\left(\sum_{j=1}^k\left(u_j-u_{j-1}-V_j(u_{j-1},u_j)\right)\geq\lambda^U_t
t\right)\\
&&\leq E\left[\prod_{j=1}^k e^{-\theta
V_j(u_{j-1},u_j)}\right]e^{\theta\left(1-\lambda^U_t\right) t}~,
\end{eqnarray*}
after using again the Chernoff bound.

At this point we rearrange the terms in the product as in
Eq.~(\ref{eq:arrange}) and proceed as for the lower bound using
H\"{o}lder's inequality first and then the independence property
of $V_j$'s over non-overlapping intervals. We immediately get
\begin{eqnarray*}
\P_k\left(D(t)\geq \lambda^U_t t\right)\leq
\left(b_{-\gamma\theta}^t\right)^{\frac{1}{\gamma}}e^{\theta\left(1-\lambda^U_t\right)
t}~,
\end{eqnarray*}
using the Laplace transform from Section~\ref{sec:tools}.

Finally, for some fixed violation probability $\eps$, the lower
bound $\lambda^L_t$ and the upper bound $\lambda^U_t$ follow by
the change of probability measure $\P=\sum_k\tilde{\pi}_k\P_k$,
which completes the proof.~\hfill $\Box$

\medskip

As mentioned in the description of the network model, the analysis
does not need the distribution of the number of nodes inside the
intersections of different ISes. The reason is that the expansion
of the virtual processes in Eq.~(\ref{eq:derive2e3}) is done over
non-overlapping intervals, whereas the virtual processes have
independent increments. The fact of non-overlapping intervals is
directly related to an essential property of the $(\min,+)$ e2e
convolution formula from Eq.~(\ref{eq:e2esc}), which expands over
such intervals. For instance, if $K=2$, then the e2e convolution
expands as
\begin{equation*}
S_1\conv S_2(s,t)=\inf_{s\leq u\leq
t}\left\{S_1(s,u)+S_2(u,t)\right\}~\forall s\leq t~.
\end{equation*}
The practical interpretation of non-overlapping intervals is that,
at some hop, the transmission of a bit is influenced by the
randomness in the network model on some time interval (say
$[s,u]$, where $s$ and $u$ are r.v.'s). At the next hop, however,
the transmission of the \textit{same} bit, is influenced by
randomness over an interval starting at $u$. If the underlying
stochastic processes had memory, then the last statement would be
false, i.e., the latter transmission can be very much influenced
by the whole past history; however, recall that in our setting,
the virtual interfering processes which incorporate the whole
randomness in the network model (see also
Eq.~(\ref{eq:derive2e3})), have independent increments. The
decoupling of the time scales at different hops for the
\textit{same} bit is an essential property of the convolution
formula.

We also remark that the major challenge of dealing with
\textit{any} possible dependencies amongst $N_j$'s, instrumented
by the value of the dependency parameter $\gamma$, is resolved by
the arrangement from Eq.~(\ref{eq:arrange}) and H\"{o}lder's
inequality. It is worth pointing out that  H\"{o}lder's inequality
holds for any dependency structure, for which reason we do not
enforce a particular correlation model amongst $N_j$'s.

Finally, we remark that, in the proof, the lower bound was derived
by applying H\"{o}lder's inequality and the Chernoff/union bounds.
The union bound, in particular, is not so bad if the r.v.'s
$X_j:=V_j(u_{j-1},u_{j})$ (see Eq.~(\ref{eq:derive2e3})) are
rather uncorrelated~\cite{tala96b}, which is the case in our
setting. A joint property of the three bounds is that they
generally capture the right scaling of e2e
results~\cite{BuLiCi11}, in our case of the e2e lower bounds. In
turn, the e2e upper bounds were derived with the bound
\begin{equation*}\P\left(\inf_{s}X_s\geq\sigma\right)\leq
\inf_s\P\left(X_s\geq\sigma\right)\label{eq:trivialbound}\end{equation*}
for some stochastic process $X_s$ (see Eq.~(\ref{eq:upbound2})).
This bound was frequently used in queueing analysis (see,
e.g.,~\cite{MontgomeryV96,Shroff98}), and was argued to be
reasonably accurate relative to a dominant time scale. We note
from Eq.~(\ref{eq:ub}), however, that this bound does not capture
the scaling of the e2e results in a non-asymptotic regime;
according to Eq.~(\ref{eq:ascap}), however, the bound is tight in
an asymptotic regime.

\subsection{Sensitivity to $N_j$'s}\label{sec:impactneighbors}
Here we discuss the capacity's sensitivity to the distribution of
the number of neighbors $N_j$'s (herein referred to as $N$). We
focus on the asymptotic capacity from Eq.~(\ref{eq:ascap}), and it
is thus sufficient to consider a single-hop transmission.

Let us firstly perform some preliminary calculations. Using
Jensen's inequality we get that
\begin{equation*}
\lambda=\sum_{l\geq 2}\pi_l p(1-p)^{l-1}\geq p(1-p)^{E[N]-1}~.
\end{equation*}
The maximum in the last term is attained for $p=\frac{1}{E[N]}$,
and thus $\lambda=\Omega\left(1/E[N]\right)$. For the same value
of $p$, and using an upper bound on Jensen's inequality (see
Theorem~1.2 in~Simic~\cite{Simic09}), we get
$\lambda={\mathcal{O}}\left(1/E[N]\right)$, and therefore the
asymptotic capacity scales as
\begin{equation}
\lambda=\Theta\left(\frac{1}{E[N]}\right)~.\label{eq:scaling1}
\end{equation}
This scaling also holds in the case of a static network in which
$N$ is a constant, i.e., $\P\left(N\neq E[N]\right)=0$.

An improved scaling law can be obtained by assuming that, on every
sample-path $\omega$, all the nodes are aware of their densities
$N_{\omega}$, and set their transmission probabilities to
$p_{\omega}=\frac{1}{N_{\omega}}$; this probability is now a
\textit{random measure}. A lower bound on the asymptotic capacity
is
\begin{eqnarray*}
\lambda^{\textrm{}}=
\sum_{l\geq2}\pi_l\frac{1}{l}\left(1-\frac{1}{l}\right)^{l-1}
\geq\sum_{l\geq2}\pi_l\frac{1}{le}\notag=\frac{1}{e}E\left[\frac{1}{N}\right]~,\label{eq:mobgd}
\end{eqnarray*}
after using
$\left(1-\frac{1}{l}\right)^{l-1}\geq\lim_{n\rightarrow\infty}\left(1-\frac{1}{n}\right)^{n-1}=\frac{1}{e}$.
In turn, an upper bound is
\begin{eqnarray*}
\lambda^{\textrm{}}=
\sum_{l\geq2}\pi_l\frac{1}{l}\left(1-\frac{1}{l}\right)^{l-1}
\leq\sum_{l\geq2}\pi_l\frac{1}{l}\notag=E\left[\frac{1}{N}\right]~.
\end{eqnarray*}

Therefore, with the sample-path neighbor-aware probabilities
assumption, the asymptotic capacity scales as
\begin{equation}
\lambda=\Theta\left(E\left[\frac{1}{N}\right]\right)~.\label{eq:scaling2}
\end{equation}
The same scaling is achieved by an ideal distributed scheduling
mechanism, such as TDMA (which can be modelled as $X_l(t)=1$ if
$t\mod N=l-1$, and $X_l(t)=0$ otherwise, in Eq.~(\ref{eq:vip})) or
by 802.11 DCF, since the transmission probabilities $p$ implicitly
scale as $\Theta\left(\frac{1}{N}\right)$
(Bianchi~\cite{Bianchi00}). Therefore, CSMA and TDMA networks
achieve the same scaling, up to the assumptions discussed in
Section~\ref{sec:model} on CSMA; this result was previously shown
to hold in the particular case of binomial node densities (i.e., a
underlying uniform geometry) and an idealized CSMA model~(see
Chau~\et~\cite{Chau09}).

Inspecting the two scalings from Eqs.~(\ref{eq:scaling1}) and
(\ref{eq:scaling2}), with Jensen's inequality
$E[N]E\left[\frac{1}{N}\right]\geq1$, reveals that the latter is
asymptotically bigger. Therefore, the capacity in a random network
with neighbor-aware transmission probabilities is asymptotically
bigger than the capacity of a random network with a fixed value
for $p$, or a static network with optimally adjusted $p$.

The discrepancy between the two scaling laws raises the
interesting question on the gain-maximizing distribution of $N$
which maximizes the capacity from Eq.~(\ref{eq:scaling2}). To
prevent trivial scenarios, such as when there are only two nodes
in the network ($\pi_2=1$), we look for the distribution of $N$
relative to the normalized, or aligned, static scenario with a
constant number $E[N]$ of nodes (note that the random and static
scenarios are aligned in that the number of nodes are identical,
on average). We are thus interested in maximizing the normalized
randomness gain
\begin{equation}
\argmax_{N}\frac{\textrm{Eq.~(\ref{eq:scaling2})}}{\textrm{Eq.~(\ref{eq:scaling1})}}=\argmax_{N}E[N]E\left[\frac{1}{N}\right]~,\label{eq:gain}
\end{equation}
subject to the sample-path node-aware probabilities assumption.
The next theorem provides the solution.

\begin{theorem}{({\sc{Gain-Maximizing Distrib. in
Eq.~(\ref{eq:gain})}})}\label{cr:optimal}\newline Denote
$n=n_{\textrm{max}}$ (the maximum value of $N$). Then
Eq.~(\ref{eq:gain}) is maximized by the distribution
\begin{equation}
\pi_2=\frac{n-E[N]}{n-2},~\pi_{n}=\frac{E[N]-2}{n-2},~\pi_i=0~(i=1,3,\dots,n-1)\label{eq:optd2}
\end{equation}
when both $n$ and $E[N]$ are fixed.
\end{theorem}

For the proof see the Appendix. The intuition behind the bimodal
distribution is that the increase rate in capacity by lowering the
number of neighbors is larger than the decrease rate in capacity
by increasing the number of neighbors. Note also that the
distribution maximizes not only Eq.~(\ref{eq:gain}), but also the
capacity when both $n$ and $E[N]$ are fixed, under the sample-path
node-aware probabilities assumption.

Table~\ref{tb:gain} illustrates the scaling of
$E[N]E\left[\frac{1}{N}\right]$ from Eq.~(\ref{eq:gain}), for
various distributions $\pi_l=\P(N=l)$. $\kappa$'s are
normalization constants, $r=1-q$ for the binomial distribution
(Bnom.), and $0<a<1$ for the geometric distribution (Geom.). There
are two versions of harmonic (Harm.), heavy-tailed (Hv.tld.),
subexponential (Sbexp.), and geometric distributions, depending on
the hops' density; for instance, row four models a sparse
situation with higher densities assigned to smaller number of
nodes, whereas row five models the opposite situation.

\begin{table}[t]
\begin{center}
\begin{tabular}{|c| c | c | c | c |}
\hline
Dist.&$\pi_l$& $E[N]$& $E\left[\frac{1}{N}\right]$&Gain \\
\hline \hline
G-M&(\ref{eq:optd2})& $\Theta(n)$& $\Theta(1)$&$\Theta(n)$\\
\hline
Unif.&$\frac{1}{n}$& $\Theta(n)$& $\Theta\left(\frac{\log n}{n}\right)$&$\Theta(\log n)$\\
\hline
Bnom.&$\binom{n}{l}(\frac{r}{q})^l q^n$& $\Theta\left(n\right)$& $\Theta\left(\frac{1}{n}\right)$&$\Theta\left(1\right)$\\
\hline
Harm.&$\frac{\kappa}{l\log n}$& $\Theta\left(\frac{n}{\log n}\right)$& $\Theta\left(\frac{1}{\log n}\right)$&$\Theta\left(\frac{n}{\log^2 n}\right)$\\
\hline
Harm.&$\frac{\kappa/\log n}{(n-l)}$& $\Theta\left(n\right)$& $\Theta\left(\frac{1}{n}\right)$&$\Theta\left(1\right)$\\
\hline
Hv.tld.&$\frac{\kappa}{l^2}$& $\Theta\left(\log n\right)$& $\Theta\left(1\right)$&$\Theta\left(\log n\right)$\\
\hline
Hv.tld.&$\frac{\kappa}{(n-l)^2}$& $\Theta\left(n\right)$& $\Theta\left(\frac{1}{n}\right)$&$\Theta\left(1\right)$\\
\hline
Sbexp.&$\frac{\kappa}{l^3}$& $\Theta\left(1\right)$& $\Theta\left(1\right)$&$\Theta\left(1\right)$\\
\hline
Sbexp.&$\frac{\kappa}{(n-l)^3}$& $\Theta\left(n\right)$& $\Theta\left(\frac{1}{n}\right)$&$\Theta\left(1\right)$\\
\hline
Geom.&$\kappa a^l$& $\Theta\left(1\right)$& $\Theta\left(1\right)$&$\Theta\left(1\right)$\\
\hline
Geom.&$\kappa a^{n-l}$& $\Theta\left(n\right)$& $\Theta\left(\frac{1}{n}\right)$&$\Theta\left(1\right)$\\
\hline
\end{tabular}
\end{center}
\caption{Capacity randomness gains for various distributions
$\mbox{\boldmath${\pi}$}=(\pi_1,\dots,\pi_n)$; $n$ is the maximum
value of $N$.} \label{tb:gain}
\end{table}

The reported gain in the last column is relative to the
expectation $E[N]$. Note that the gain-maximizing (G-M)
distribution from Eq.~(\ref{eq:optd2}) with $E[N]=\Theta(n)$ and
the heavy-tailed distribution modelling sparse situations achieve
the maximum relative gain, further supporting the observation
after Theorem~\ref{cr:optimal}. The uniform (Unif.) distribution
achieves the same gain $\Theta(\log n)$ as the first heavy-tailed
distribution, but that is relative to an asymptotically larger
expected number of nodes, i.e., $\Theta(n)$ vs. $\Theta(\log n)$.

Summarizing the results, we conclude that the asymptotic (in the
time scale) per-flow capacity is fundamentally influenced by
randomness in network geometry,~\textit{but only} under the
sample-path node-aware probabilities assumption. This assumption
requires explicit overhead for both TDMA and Aloha schemes, and it
is implicitly satisfied by 802.11 DCF since the nodes'
transmission probabilities are $p_{\omega}=\Theta(1/N_{\omega})$
in steady-state (Bianchi~\cite{Bianchi00}). Similar conclusions
can be drawn on the non-asymptotic capacity as well, since the
bounds from Eqs.~(\ref{eq:lb}) and (\ref{eq:ub}), with
$\theta=\Theta\left(t^{-\zeta}\right)$, deviate from
Eq.~(\ref{eq:ascap}) by constant terms depending on the time scale
and/or the number of hops.

Let us clarify the apparent contradiction between the above
observation that ``randomness increases the per-flow capacity" and
the folk principle that ``\textit{determinism minimizes the
queue}" from queueing theory (Humblet~\cite{Humblet82}) (which
agrees in particular with the fact that randomness decreases the
network capacity~\cite{Gupta00}). The reason is that our network
model is slightly different than the one from~\cite{Gupta00},
specifically by fixing both the source and the destination and
letting the rest be random. Recall that our network model is
deliberately tailored to directly study the per-flow rather than
the network capacity.

\subsection{Sensitivity to $K$}\label{sec:senK}
Here we analyze the role of the distribution of the number of hops
$K$ on the lower bound of the non-asymptotic capacity from
Eq.~(\ref{eq:lb}); as already pointed out, the other capacity
results (i.e., the upper bound from Eq.~(\ref{eq:ub}) and the
asymptotic one from Eq.~(\ref{eq:ascap})) are invariant to $K$.
Despite this apparent incompleteness (we only analyze the lower
bounds), we conjecture that the obtained scaling laws herein hold
for upper bounds as well, given the previous observation that the
lower bound captures the right scaling in $K$.

Analyzing the scaling law in $K$ in Eq.~(\ref{eq:lb}) yields a
trivial result, i.e., $\Theta(1)$, since the limit in $t$ must be
taken simultaneously (recall that $t\geq k_{max}$ in
Theorem~\ref{th:e2ecap}). Informally, note that if
$t=\Theta\left(E[K]^{1-\zeta}\right)$, for $\zeta>0$, then
$\lambda_t^L=0$ because there are insufficient time slots to carry
packets from the source to the destination. To get more
interesting results, let us properly scale
$t=\Theta\left(E[K]^{1+\zeta}\right)$ and $\gamma=\Theta(1)$, for
some large $E[K]$. Then, the lower bound $\lambda_t^L$ decays as
\begin{equation*}
\lambda_t^L=\Omega\left(-\log{E[K]}\right)~,
\end{equation*}
in the case of a static network with a fixed number $E[K]$ of
hops. In turn, in the case of a network with a random number $K$
of hops, $\lambda_t^L$ decays as
\begin{equation*}
\lambda_t^L=\Omega\left(-E\left[\log{K}\right]\right)~.
\end{equation*}

Jensen's inequality ($E\left[\log{K}\right]\leq \log E[K]$)
implies that the lower bound on capacity in a random scenario is
asymptotically bigger than in a static scenario. As in the
previous subsection, this discrepancy raises the problem of the
gain-maximizing distribution of $K$ which maximizes the randomness
gain, defined here as
\begin{equation}
\argmax_{K}\left(\log{E[K]}-E\left[\log{K}\right]\right)~.\label{eq:gainK}
\end{equation}
Before we provide the answer, in the next theorem, let us remark
that the randomness gain is now defined in terms of a
\textit{difference}, and not of a \textit{ratio} as in
Eq.~(\ref{eq:gain}). The reason stands in the contribution of $K$
and $N$ to the e2e capacity: the former has an additive effect
(i.e., it affects the \textit{cumulative throughput}), whereas the
latter has a multiplicative effect (i.e., it affects the
\textit{throughput rate}).

\begin{theorem}{({\sc{Gain-Maximizing Distrib. in
Eq.~(\ref{eq:gainK})}})}\label{cr:optimalK}\newline Denote
$k=k_{\textrm{max}}$ (the maximum value of $K$). Then
Eq.~(\ref{eq:gainK}) is maximized by the distribution
\begin{equation}
\tilde{\pi}_1=\frac{k-E[K]}{k-1},~\tilde{\pi}_k=\frac{E[K]-1}{k-1},~\tilde{\pi}_i=0~(i\neq
1,k)\label{eq:optd2K}
\end{equation}
when both $k$ and $E[K]$ are fixed.
\end{theorem}
The proof is similar to the proof of Theorem~\ref{cr:optimal} and
it is omitted.

\medskip

Similar as in Theorem~\ref{cr:optimal}, the intuition for the
bimodal distribution is that the rate at which capacity increases
by lowering the number of hops is bigger than the rate at which
capacity decreases by increasing the number of hops. Note that the
distribution from Eq.~(\ref{eq:optd2K}) maximizes not only the
randomness gain from Eq.~(\ref{eq:gainK}) but also the capacity
(its lower bound) when both $k$ and $E[K]$ are fixed. Also, note
that the randomness gain of the distribution from
Eq.~(\ref{eq:optd2K}) can be asymptotically larger than a constant
(e.g., $\Theta\left(\log{\log{k}}\right)$ vs.
$\Theta\left(\frac{\log^2{k}}{k}\right)$ when
$E[K]=\Theta(\log{k})$).

As far as other distributions are concerned, recall that in the
case of the Uniform distribution for the number of neighbors,
Table~\ref{tb:gain} reports a randomness gain of
$\Theta\left(\log{n}\right)$. In contrast, there is no randomness
gain in the case of the Uniform distribution for the number of
hops. The reason is that while increasing the number of neighbors
has a pronounced effect on capacity, increasing the number of hops
has a much more moderate effect due to spatial reuse. This also
indicates that spatial correlations are much less sensitive to
randomness, as opposed to temporal correlations as observed in the
previous subsection. As another closely related example, the first
heavy-tailed distribution for $N$ from Table~\ref{tb:gain} yields
a $\Theta\left(\log{n}\right)$ gain in Eq.~(\ref{eq:gain}). In
turn, using the convergence of the series
$\sum_{l=1}^k\frac{\log{l}}{l^2}$, the same heavy-tailed
distribution for $K$ has the gain $\Theta\left(\log\log k\right)$
in Eq.~(\ref{eq:gainK}).

The above example leads us to speculate that for specific
distributions of $N$ and $K$, the (per-flow) capacity gain due to
randomness in the number of hops is the logarithm of the capacity
gain due to randomness in nodes' densities.

\subsection{Sensitivity to $\gamma$}\label{sec:sengamma}
Here we briefly investigate the role of the dependency parameter
$\gamma$ on the non-asymptotic capacity.

Note firstly that taking a limit in $\gamma$ would require taking
limits in both $E[K]$ and $t$. As in the previous subsection, let
$t=\Theta\left(E[K]^{1+\zeta}\right)$ for some large $E[K]$, but
take now $\gamma=\Theta(E[K])$ which models a high degree of
dependencies amongst $N_j$'s (including the worst-case when all
hops' densities are correlated to each other). Then, the lower
bound decays as
\begin{equation}
\lambda_t^L=\Omega\left(-\log{E[K]}\right)~,\label{eq:logdecga}
\end{equation}
i.e., the logarithm is the price for assuming a high degree of
spatial dependencies in the network. See
also~Burchard~\et~\cite{BuLiCi11} where the same logarithmic
factor is the additional decay on e2e delays (in a $\Theta(\cdot)$
sense) in networks with Markovian type of traffic, due to spatial
dependencies; recall Item O4 from the Introduction.

Wrapping up, the main observations from this section are
summarized in Items O1-O4 from the Introduction. These
observations raise an interesting tradeoff between per-flow
fairness/delay vs. capacity metrics, which has been addressed at
the network level in the particular case of uniform geometries
(see Grossglauser and Tse~\cite{Grossglauser02}, Neely and
Modiano~\cite{NeelyM05a}, and Sharma~\et~\cite{Sharma07}). We
believe that a further understanding of this tradeoff at the
per-flow level, in networks with general geometries, can inspire
distributed network algorithms emulating randomness, in order to
provide differentiated per-flow services while maintaining a
certain level of performance at the network level. More concisely,
how could one leverage the randomness in the network geometry,
given its advantageous impact on per-flow capacity?

\section{Conclusions}\label{sec:conclusions}
We have derived closed-form per-flow capacity results, in terms of
both upper and lower (non-)asymptotic bounds, on a multi-hop path
with a fixed source-destination pair. The key aspect of these
results is that they apply to networks with \textit{general random
geometries} and various degrees of spatial correlations. By
exploiting the simple analytical forms of the obtained results, we
have quantified the beneficial impact of randomness in geometry on
the per-flow capacity metric. In particular, we have shown that
different distributions of hops' densities with normalized
averages can lead to gaps in capacity scaling laws as large as
${\Theta}(n)$. We have identified a logarithmic detrimental factor
of spatial correlations, and we have further observed a
logarithmic relationship between spatial and temporal
correlations.

Beyond the intuitively obvious message that randomness matters,
this paper strives to communicate how does randomness
\textit{quantitatively} matter by analyzing a broad range of
distributions. The collected observations jointly raise the
awareness that the restriction to the widely-adopted uniform
geometry model can be quite misleading. Moreover, these
observations open the conceivably practical and yet challenging
research problem of increasing per-flow capacity by leveraging the
randomness in network geometry.


\bibliographystyle{abbrv}
\bibliography{../../stat}
\appendix

{\sc Proof of Theorem~\ref{cr:optimal}.}~According to
Eq.~(\ref{eq:mobgd}) we have to solve a linear program with the
objective function
\begin{equation*}
\max_{\pi_i}\sum_{i\geq2}\frac{1}{i}\pi_i
\end{equation*}
subject to the constraints
\begin{equation*}
\sum_{i\geq1} i\pi_i=m,~\sum_{i\geq1}\pi_i=1,~\pi_i\geq0~.
\end{equation*}

Recall that $\pi_1=0$ (there are at least two nodes). Select some
$\pi_i$'s satisfying the above constraints and assume that there
exists $i\in\left\{3,\dots,n-1\right\}$ such that $\pi_i\neq0$.
The idea is to appropriately redistribute the entire value of
$\pi_i$ to the extremes $\pi_2$ and $\pi_n$, and then show that
the new distribution increases the objective function.

Let $x=\frac{n-i}{n-2}\pi_i$ and consider the new distribution
$\pi'_i$ similar to $\pi_i$ except for
\begin{equation*}
\pi'_2=\pi_2+x,~\pi'_i=0,~\pi'_n=\pi_n+\pi_i-x~.
\end{equation*}
One can check that the redistribution of $\pi_i$ preserves the
mean $m$. It remains to prove that
\begin{eqnarray*}
\frac{1}{2}\pi_2+\frac{1}{i}\pi_i+\frac{1}{n}\pi_n\leq\frac{1}{2}(\pi_2+x)+\frac{1}{n}\left(\pi_n+\pi_i-x\right)~.
\end{eqnarray*}
By rearranging terms the inequality reduces to $i\geq2$ which is
true.

One can repeat the above procedures for all $i$ in the set
$\left\{3,\dots,n-1\right\}$ satisfying $\pi_i\neq0$, and reduce
the initial linear program to
\begin{equation*}
\max_{\pi_2,\pi_n}\left\{\frac{1}{2}\pi_2+\frac{1}{n}\pi_n\right\}~,
\end{equation*}
subject to the constraints
\begin{equation*}
2\pi_2+n\pi_n=m,~\pi_2+\pi_n=1,~\pi_2\geq0,~\pi_n\geq0~.
\end{equation*}
The solution is given by the choice from
Eq.~(\ref{eq:optd2}).\hfill $\Box$

\end{document}